\documentclass{elsart}

 \usepackage{epsfig}

\usepackage{amssymb}

\begin{document}

\begin{frontmatter}



\title{Chargino Production and Decay in Photon-Photon-Collisions}


\author{T. Mayer\thanksref{mail}},
\author{H. Fraas\thanksref{mail2}}
\thanks[mail]{e-mail:mayer@physik.uni-wuerzburg.de}
\thanks[mail2]{e-mail:fraas@physik.uni-wuerzburg.de}

\address{Institut f\"{u}r Theoretische Physik, Universit\"at W\"urzburg,\\ Am
  Hubland, D-97074 W\"urzburg, Germany}

\begin{abstract}
We study the production and leptonic decay of charginos in collisions
of polarized photon beams including the
complete spin correlations. The photons can be generated by Compton
backscattering of polarized laser pulses off a polarized electron
beam. Since the production process is determined alone by the
electromagnetic coupling of the charginos this process allows to study
their decay dynamics. The cross section and the
forward-backward asymmetry of the decay lepton are very sensitive to
the gaugino mass parameter $M_1$ and to the sneutrino mass without any 
ambiguities.
\end{abstract}


\end{frontmatter}

\section{Introduction}
The search for supersymmetric (SUSY) particles and the determination
of their properties is one of the main goals of
a future $e^+e^-$ Linear Collider in the energy range between 500 GeV
and 1000 GeV. Particularly interesting will be the experimental study
of charginos. They will be produced with comparably large cross
sections and the analysis of their production and decay will allow to
measure the parameters of the underlying supersymmetric model. 
In the Minimal Supersymmetric Standard Model (MSSM) \cite{haberk} with 
conserved R parity the chargino masses and couplings are determined by the
SUSY parameters $M_2$, $\mu$ and $\tan\beta$. The neutralino
properties and thus the chargino decay into the lightest neutralino
$\tilde{\chi}^0_1$, which is assumed to be the lightest SUSY particle
(LSP), depend in addition on the
gaugino mass parameter $M_1$. Recently a method has been proposed to
separate the chargino production process from the chargino decay and
to determine the SUSY parameters $M_2$, $\mu$ and $\tan\beta$
independently of the neutralino sector \cite{choi}. 

Besides the
$e^+e^-$ option also the $\gamma\gamma$ mode of a Linear Collider can be
realized with high luminosity polarized photon beams obtained by
Compton backscattering of laser pulses off the electron beam. In this
contribution we study chargino pair production
$\gamma\gamma\to\tilde{\chi}^+_i\tilde{\chi}^-_i$ $(i=1,2)$ and for
the case of the lighter charginos the subsequent
leptonic decay $\tilde{\chi}^+_1\to\tilde{\chi}^0_1e^+\nu_e$. The
chargino exchange in the t- and u-channel of the production process is 
determined by the electromagnetic coupling of the chargino and, apart
from the chargino mass, completely independent of the SUSY
parameters. Therefore the combined process of production and decay
allows to study the decay mechanism separated from the model
independent production mechanism. The unpolarized production cross
section for the charginos is not
suppressed by destructive interference effects \cite{bartl} and larger
than for chargino production in 
$e^+e^-$ annihilation by a factor of 2.4 at $\sqrt{s}=500$ GeV and
even by a factor of 9 at $\sqrt{s}=1000$ GeV. By appropriate
choice of the polarization of the photon beams it can be enhanced. As a
further advantage the polarization of the chargino is solely 
determined by the polarization of the photon beams and can be modified 
by changing the polarization of the laser beam and the converted
electron beam.

\section{Cross Sections and Decay Angular Distributions}
The helicity \sloppy amplitudes for the pure electromagnetic production
process $\gamma(\alpha)\gamma(\beta)\to
\tilde{\chi}_i^+(\lambda_i)\tilde{\chi}_i^-(\lambda_j)$ proceeding
via chargino exchange in the t- and u-channel are
denoted by $T^{\lambda_i\lambda_j}_{P,\alpha \beta}$. Here
$\alpha,\beta$ are the helicities of the photons 
and $\lambda_i,\lambda_j$ the helicities of the charginos. The direct
leptonic decay
$\tilde{\chi}^\pm_i\to\tilde{\chi}^0_1e^\pm\stackrel{(-)}{\nu_e}$ proceeds via
$W^\pm,\tilde{e}_L$ and 
$\tilde{\nu}_e$ exchange with helicity amplitudes $T_D^{\lambda_i}$
($T_D^{\lambda_j}$). The amplitude squared of the combined process of
production and decay is (summed over chargino helicities)
\begin{equation}
|T_{\alpha\beta}|^2=|\Delta(\tilde{\chi}_i^+)|^2|\Delta(\tilde{\chi}_i^-)|^2\rho^{\lambda_i\lambda_j,\lambda^\prime_i\lambda^\prime_j}_{P,\alpha\beta}\rho_D^{\lambda_i^\prime\lambda_i}\rho_D^{\lambda^\prime_j\lambda_j}.
\end{equation}
It is composed of the (unnormalized) spin density production matrix
\begin{equation}
\rho^{\lambda_i\lambda_j,\lambda^\prime_i\lambda^\prime_j}_{P,\alpha\beta}=T^{\lambda_i\lambda_j}_{P,\alpha \beta}\;T^{\lambda^\prime_i\lambda^\prime_j\ast}_{P,\alpha \beta},
\end{equation}
which is, apart from the chargino mass, independent of the SUSY
parameters, the decay matrices
\begin{equation}
\rho_D^{\lambda_i^\prime\lambda_i}=T_D^{\lambda_i}T_D^{\lambda^\prime_i\ast} 
\qquad\rho_D^{\lambda_j^\prime\lambda_j}=T_D^{\lambda_j}T_D^{\lambda^\prime_j\ast}
\end{equation}
and the propagator
$\Delta(\tilde{\chi}_i)=\frac{1}{p_i^2-m_i^2+im_i\Gamma_i}$. Here
$p_i^2,m_i$ and $\Gamma_i$ are the four-momentum squared, mass and
width of $\tilde{\chi}_i^\pm$. For this propagator we use the narrow
width approximation. For further details of this spin formalism
including full spin correlations between production and decay see
\cite{gudi,gudineut}. The production density matrix
$\rho^{\lambda_i\lambda_j,\lambda^\prime_i\lambda^\prime_j}_{P,\alpha\beta}$ 
for polarized photons will be given in a forthcoming paper. The decay
density matrix can be found in \cite{gudi}. The photon beams are
produced by Compton 
backscattering of circularly polarized laser photons off longitudinal
polarized electrons. The energy spectrum and the mean helicity of the
high energy photons sensitively depend on the polarization of the
laserphotons and of the converted electrons and are given in
\cite{ginzburg}. To
obtain the cross sections and the angular distributions of the decay
electrons in the laboratory frame (ee-cms) one has to convolute the
cross section in the $\gamma\gamma$-cms with the energy distribution
and the mean helicity of the backscattered photon beams
\cite{kon,hesselb}. In the next section we give numerical results for
the production and subsequent leptonic decay
$\tilde{\chi}^+_1\to\tilde{\chi}^0_1e^+\nu_e$ of the lighter
chargino and for the forward-backward asymmetry
\begin{equation}\label{afb}
A_{\mathrm{FB}}=\frac{\sigma(\cos\theta>0)-\sigma(\cos\theta<0)}{\sigma(\cos\theta>0)+\sigma(\cos\theta<0)}
\end{equation}
of the decay positron in the laboratory frame. In equ. (\ref{afb}) $\theta$
describes the angle between the decay positron and the converted electron beam.

\section{Numerical Results}
For the numerical analysis we fix the MSSM parameters
\begin{equation}
M_2=152\; \mathrm{GeV},\;\; \mu=316\; \mathrm{GeV},\;\; \tan\beta=3,
\end{equation}
which lead to a gaugino-like lightest chargino $\tilde{\chi}_1^\pm$ with the
mass $m_{\tilde{\chi}_1^\pm}=128$ GeV.

\begin{figure}[p]
\centering
\setlength{\unitlength}{1cm}

\begin{picture}(14,8.5)
\put(0,-3){\epsfig{file=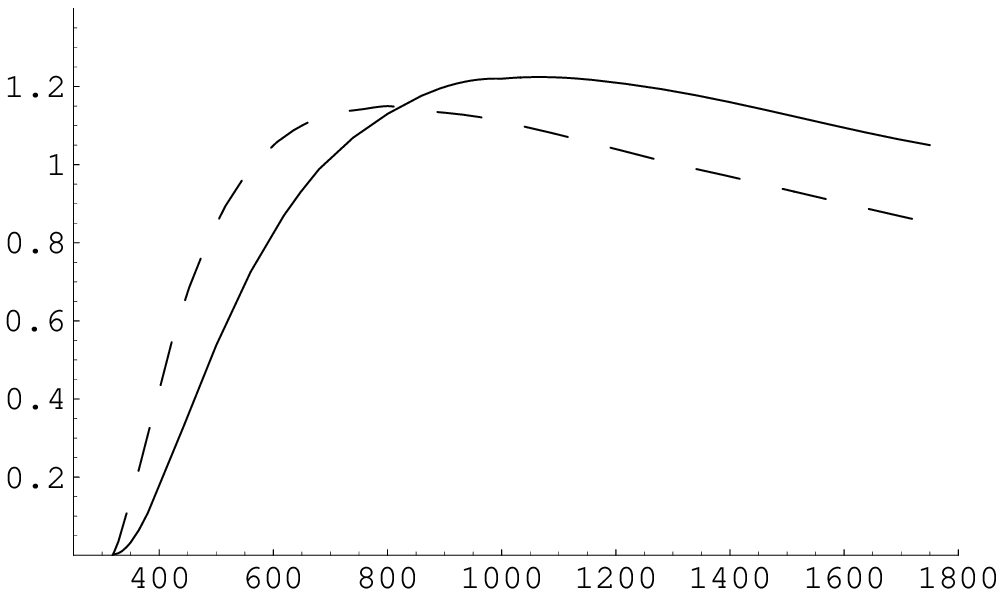,scale=1.26}}
\put(.3,7.7){\small $\sigma_{\mathrm{p}}/\mathrm{pb}$}
\put(12.6,.4){\small $\sqrt{s_{ee}}/\mathrm{GeV}$}
\end{picture}
\caption[]{\label{fig1}Production cross section 
  $\sigma_{\mathrm{p}}(\gamma\gamma\to\tilde{\chi}^+_1\tilde{\chi}^-_1)$ as a function of 
  the cms energy for $m_{\tilde{\chi}_1^\pm}=128$ GeV, $(\lambda_{k_1},\lambda_{L_1})=(1,0)$,
  $(\lambda_{k_2},\lambda_{L_2})=(1,0)$ (dashed line) and
  $(\lambda_{k_1},\lambda_{L_1})=(1,0)$,
  $(\lambda_{k_2},\lambda_{L_2})=(-1,0)$ (solid line).}
\end{figure}

In fig.~\ref{fig1} we show the energy dependence of the production
cross section
$\sigma_{\mathrm{p}}(e^+e^-\to\tilde{\chi}_1^+\tilde{\chi}_1^-)$ for
different polarization configurations. The optimal polarization
configuration depends on the beam energy. For $\sqrt{s_{ee}}=500$ GeV
one obtains the highest cross section for
$(\lambda_{k_1},\lambda_{L_1})=(1,0)$ and
$(\lambda_{k_2},\lambda_{L_2})=(1,0)$, whereas for $\sqrt{s_{ee}}>800$ GeV
the configuration $(\lambda_{k_1},\lambda_{L_1})=(1,0)$,
$(\lambda_{k_2},\lambda_{L_2})=(-1,0)$ is favoured.

\begin{figure}[t]
\centering
\setlength{\unitlength}{1cm}
\begin{picture}(14.2,9.8)
\put(0,0){\epsfig{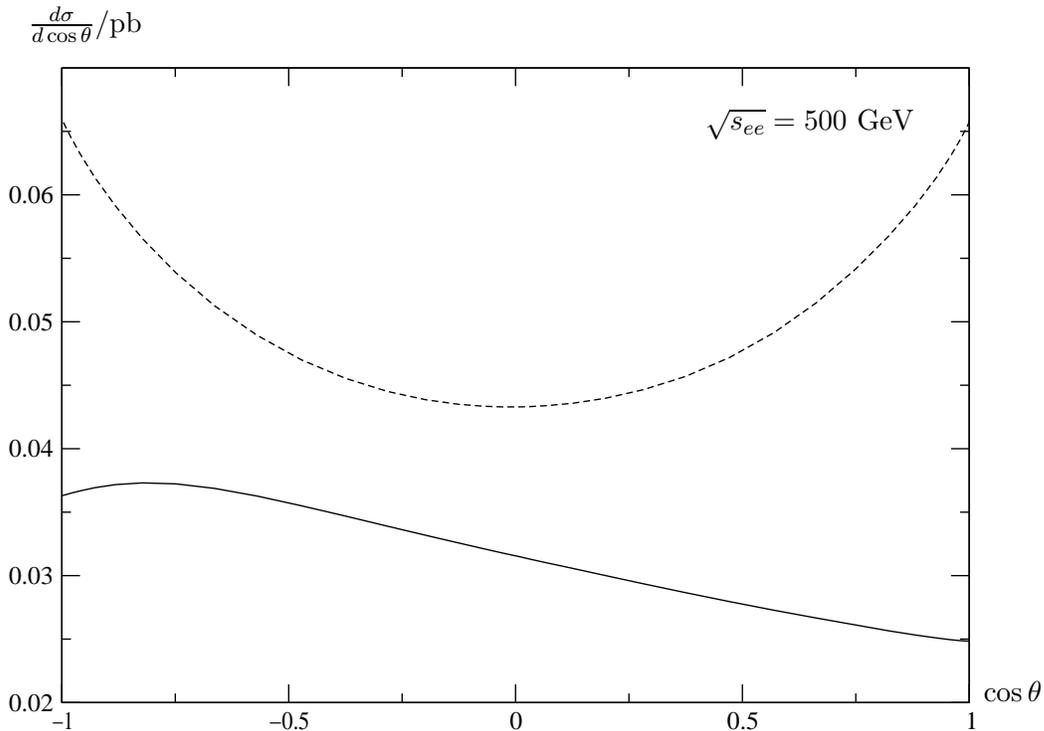}}

\put(0.3,9.3){\small $\frac{d\sigma}{d\cos\theta}/$pb}
\put(13.,0.4){\small $\cos\theta$}
\put(9.3,8){\small $\sqrt{s_{ee}}=500$ GeV}
\end{picture}

\caption[]{\label{fig2}Angular distribution of the decay positron in
  $\gamma\gamma\to\tilde{\chi}^+_1\tilde{\chi}^-_1$,
  $\tilde{\chi}^+_1\to\tilde{\chi}^0_1e^+\nu_e$ at
  $\sqrt{s_{ee}}=500$ GeV for $m_{\tilde{\nu}_e}=234$ GeV, $M_1=78.8$ GeV, 
  $(\lambda_{k_1},\lambda_{L_1})=(1,0)$,
  $(\lambda_{k_2},\lambda_{L_2})=(1,0)$ (dashed line) and
 $(\lambda_{k_1},\lambda_{L_1})=(1,0)$,
  $(\lambda_{k_2},\lambda_{L_2})=(-1,0)$ (solid line).}

\end{figure}

In fig. \ref{fig2} the angular distribution of the positron from the
leptonic decay $\tilde{\chi}^+_1\to\tilde{\chi}^0_1e^+\nu_e$ is
depicted for $\sqrt{s_{ee}}=500$ GeV. It additionally depends on the
SUSY parameter $M_1$ and on the masses of the sneutrino and the left
selectron. As an example we have chosen $m_{\tilde{\nu}_e}=234$ GeV
and $M_1=78.7$ GeV according to the GUT relation
$M_1=\frac{5}{3}\tan^2\theta_WM_2$. This corresponds to the
DESY/ECFA reference scenario for the Linear Collider \cite{ambro}. The
mass of the left selectron is
determined by the SU(2$)_L$ relation \cite{martin}
\begin{equation}\label{su2}
m^2_{\tilde{e}_L}=m^2_{\tilde{\nu}_e}-m_W^2\cos2\beta. 
\end{equation}
For the polarization configuration $(\lambda_{k_1},\lambda_{L_1})=(1,0)$,
  $(\lambda_{k_2},\lambda_{L_2})=(1,0)$ the positron angular
  distribution is forward-backward symmetric, whereas for
  $(\lambda_{k_1},\lambda_{L_1})=(1,0)$ and
  $(\lambda_{k_2},\lambda_{L_2})=(-1,0)$ the backward direction is
  preferred with a rather large asymmetry $A_{\mathrm{FB}}=-11.6\%$.
Unpolarized beams lead to a vanishing forward-backward asymmetry.

\begin{figure}[h]
\centering

\setlength{\unitlength}{1cm}
\begin{picture}(13.6,9.4)

\put(0,0){\epsfig{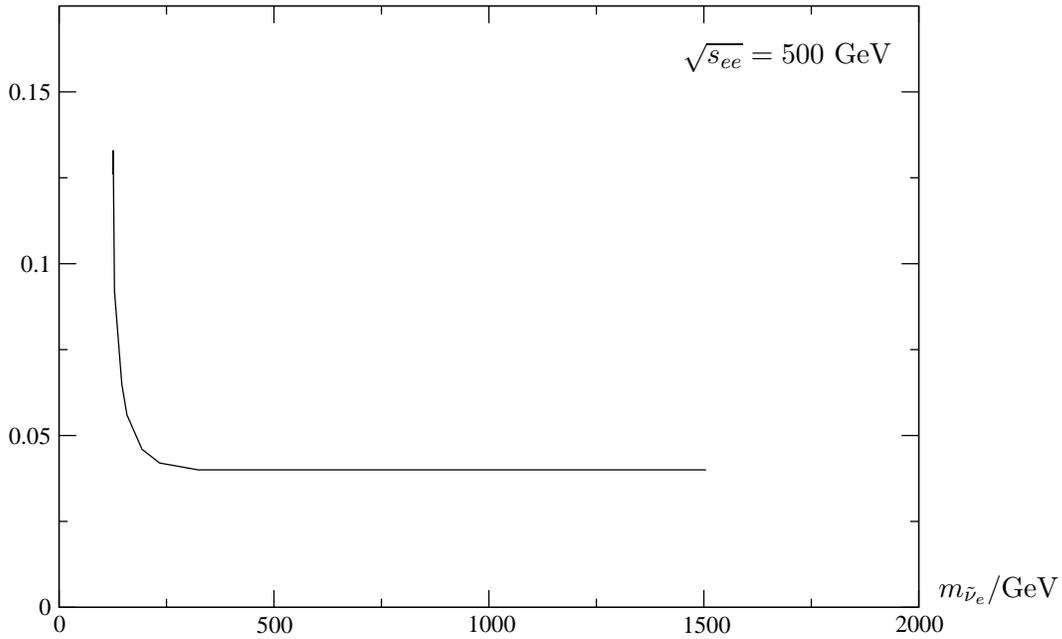}}
\put(0.4,8.9){\small$\sigma=\sigma_p\times BR(\tilde{\chi}^+_1\to\tilde{\chi}^0_1e^+\nu_e)/\mathrm{pb}$}
\put(12.4,0.5){\small$m_{\tilde{\nu}_e}/\mathrm{GeV}$}
\put(9,7.6){\small $\sqrt{s_{ee}}=500$ GeV}
\end{picture}

\caption[]{\label{fig3}Cross section for $\gamma\gamma\to\tilde{\chi}^+_1\tilde{\chi}^-_1$,
  $\tilde{\chi}^+_1\to\tilde{\chi}^0_1e^+\nu_e$ as a function of the
  sneutrino mass at $\sqrt{s_{ee}}=500$ GeV for $M_1=78.7$ GeV,
  $(\lambda_{k_1},\lambda_{L_1})=(1,0)$ and
  $(\lambda_{k_2},\lambda_{L_2})=(-1,0)$.}
\end{figure}

\begin{figure}[t]
\centering
\setlength{\unitlength}{1cm}
\begin{picture}(14,9.4)

\put(0,0){\epsfig{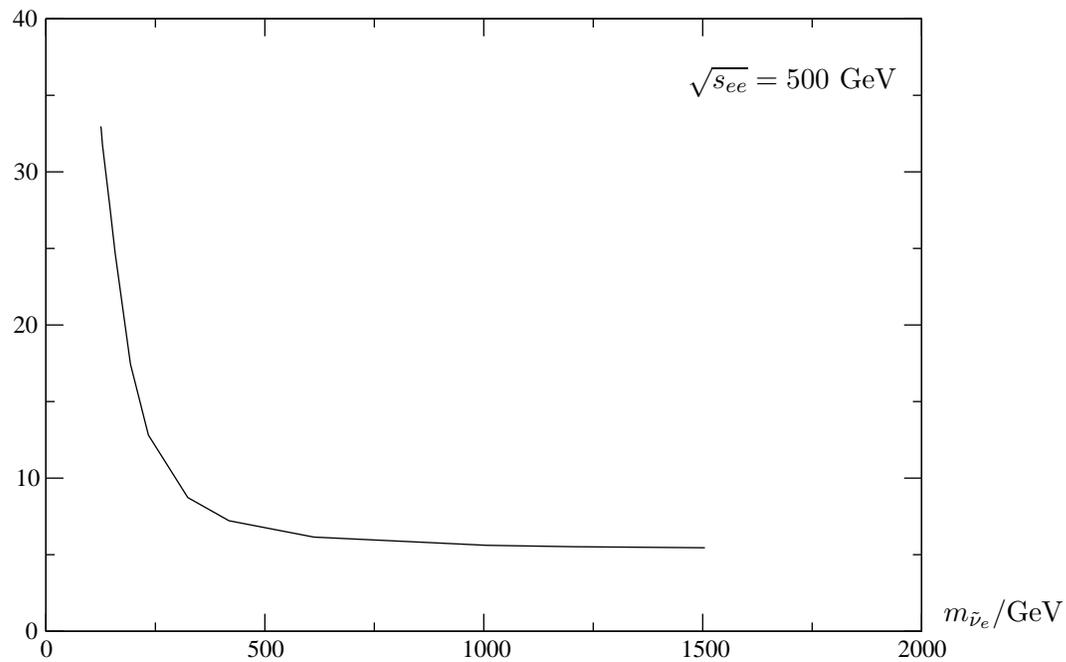}}
\put(12.4,0.5){\small$m_{\tilde{\nu}_e}/\mathrm{GeV}$}
\put(0.,9){\small$-A_{\mathrm{FB}}/\%$}
\put(9,7.6){\small $\sqrt{s_{ee}}=500$ GeV}
\end{picture}

\caption[]{\label{fig4}Forward-Backward asymmetry for
  $\gamma\gamma\to\tilde{\chi}^+_1\tilde{\chi}^-_1$,
  $\tilde{\chi}^+_1\to\tilde{\chi}^0_1e^+\nu_e$ as a function of the
  sneutrino mass at $\sqrt{s_{ee}}=500$ GeV for $M_1=78.7$ GeV,
  $(\lambda_{k_1},\lambda_{L_1})=(1,0)$ and
  $(\lambda_{k_2},\lambda_{L_2})=(-1,0)$.}
\end{figure}

Fig.~\ref{fig3} shows for $\sqrt{s_{ee}}=500$ GeV and $M_1=78.7$
GeV the $m_{\tilde{\nu}_e}$ dependence of the cross section
$\sigma=\sigma_{\mathrm{p}}
\times BR(\tilde{\chi}^+_1\to\tilde{\chi}^0_1e^+\nu_e)$ 
for beam polarizations $(\lambda_{k_1},\lambda_{L_1})=(1,0)$ and
  $(\lambda_{k_2},\lambda_{L_2})=(-1,0)$. Since the chargino
  production cross section $\sigma_{\mathrm{p}}$ is independent of the 
  SUSY parameters fig.~\ref{fig3} reflects the $m_{\tilde{\nu}_e}$
  dependence of 
  the leptonic branching ratio. The selectron mass is chosen 
  according to equ. (\ref{su2}). Fig. \ref{fig4} shows the corresponding 
  forward-backward asymmetry of the decay positron. For
  $m_{\tilde{\nu}_e}<250$ GeV the cross section shows a significant
  $m_{\tilde{\nu}_e}$ dependence whereas $A_{\mathrm{FB}}$ is
  sensitive to the sneutrino mass up to $m_{\tilde{\nu}_e}\sim400$
  GeV. Since neither the cross section nor $A_{\mathrm{FB}}$ show
  ambiguities this process should allow to determine the
  sneutrino mass in the region
  $m_{\tilde{\nu}_e}\stackrel{\textstyle<}{\sim}400$ GeV. With
  increasing $m_{\tilde{\nu}_e}$ the contributions from $\tilde{\nu}_e$
  and $\tilde{e}_L$ exchange to the chargino decay are more and more
  suppressed so that finally only the contribution from $W$ exchange
  survives.

\begin{figure}[h]
\setlength{\unitlength}{1cm}
\begin{picture}(14,9.4)

\put(0,0){\epsfig{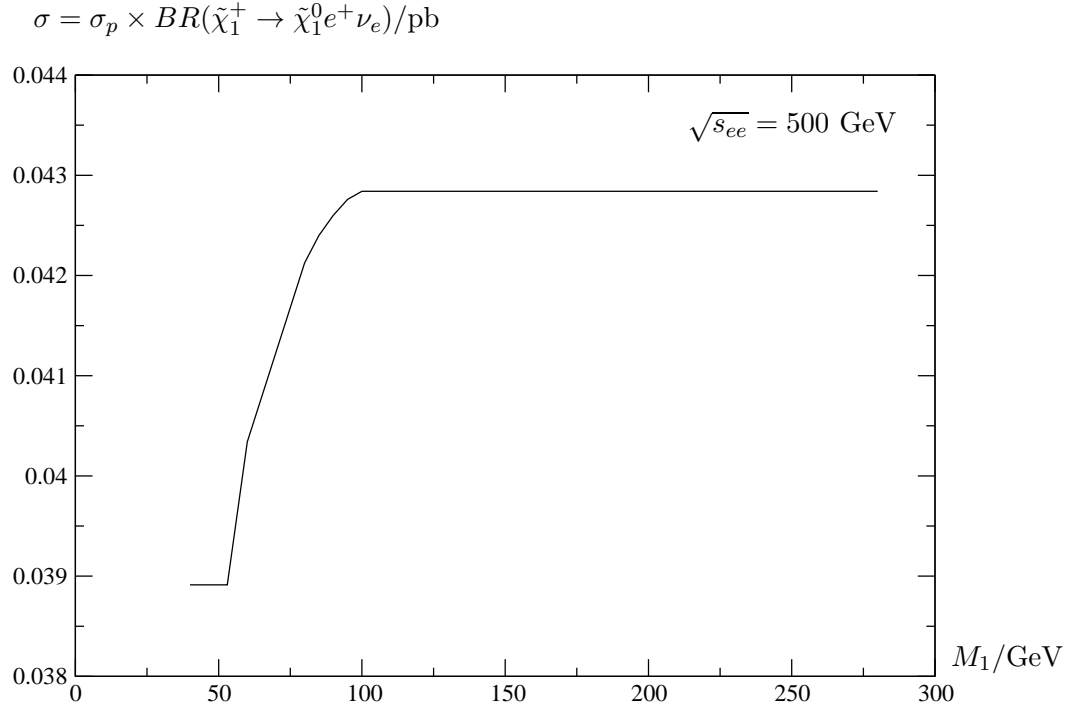}}
\put(0.3,9.){\small$\sigma=\sigma_p\times BR(\tilde{\chi}^+_1\to\tilde{\chi}^0_1e^+\nu_e)/\mathrm{pb}$}
\put(12.5,0.5){\small$M_1/\mathrm{GeV}$}
\put(9,7.6){\small $\sqrt{s_{ee}}=500$ GeV}
\end{picture}

\caption[]{\label{fig5}Cross section for $\gamma\gamma\to\tilde{\chi}^+_1\tilde{\chi}^-_1$,
  $\tilde{\chi}^+_1\to\tilde{\chi}^0_1e^+\nu_e$ as a function of the
  parameter $M_1$ at $\sqrt{s_{ee}}=500$ GeV for $m_{\tilde{\nu}_e}=234$ 
  GeV,
  $(\lambda_{k_1},\lambda_{L_1})=(1,0)$ and 
  $(\lambda_{k_2},\lambda_{L_2})=(-1,0)$.}
\end{figure}

\begin{figure}[p]
\setlength{\unitlength}{1cm}
\begin{picture}(14,9.4)

\put(0,0){\epsfig{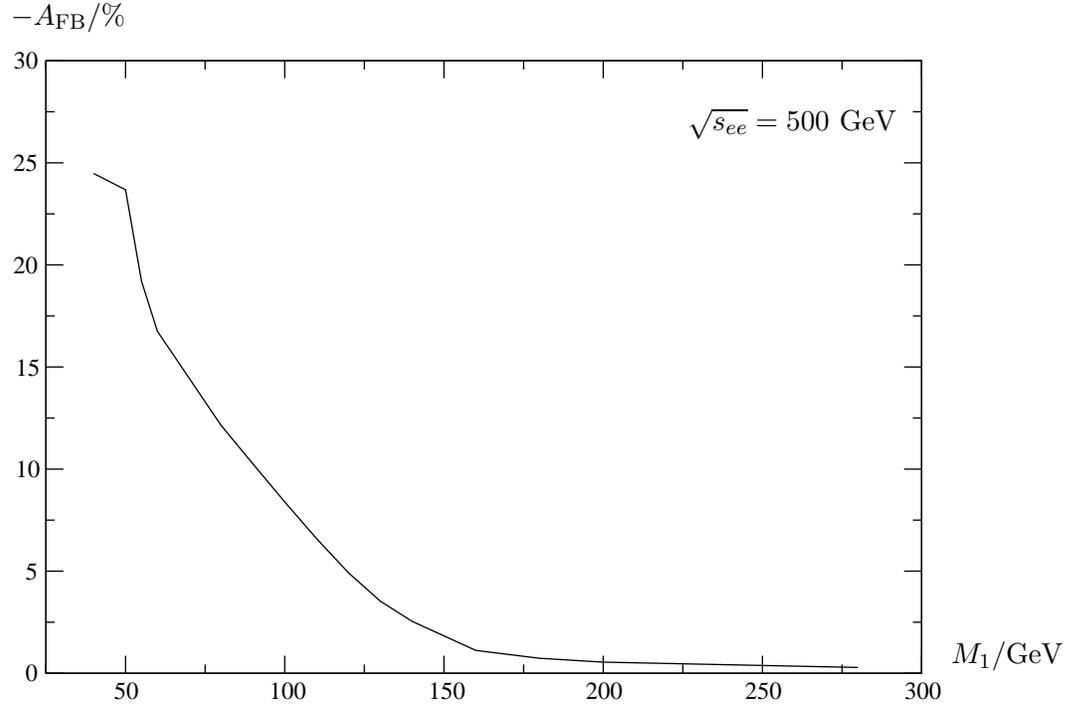}}
\put(12.5,0.5){\small$M_1/\mathrm{GeV}$}
\put(0,9.){\small$-A_{\mathrm{FB}}/\%$}
\put(9,7.6){\small $\sqrt{s_{ee}}=500$ GeV}
\end{picture}

\caption[]{\label{fig6}Forward-Backward asymmetry for
  $\gamma\gamma\to\tilde{\chi}^+_1\tilde{\chi}^-_1$,
  $\tilde{\chi}^+_1\to\tilde{\chi}^0_1e^+\nu_e$ as a function of the
  parameter $M_1$ at $\sqrt{s_{ee}}=500$ GeV for $m_{\tilde{\nu}_e}=234$ 
  GeV,
  $(\lambda_{k_1},\lambda_{L_1})=(1,0)$ and 
  $(\lambda_{k_2},\lambda_{L_2})=(-1,0)$.}
\end{figure}

Finally we give up the GUT-relation between $M_1$ and $M_2$ and show
for $\sqrt{s_{ee}}=500$ GeV in figs. \ref{fig5} and \ref{fig6},
respectively, the cross section and 
$A_{\mathrm{FB}}$ as a function of the gaugino mass parameter $M_1$ in
the region
40 GeV$<M_1<280$ GeV. The beam polarizations are again
$(\lambda_{k_1},\lambda_{L_1})=(1,0)$ and
$(\lambda_{k_2},\lambda_{L_2})=(-1,0)$. For the sneutrino mass we have 
chosen $m_{\tilde{\nu}_e}=234$ GeV and the selectron mass is
$m_{\tilde{e}_L}=245$ GeV, corresponding to equ. (\ref{su2}). The
decay $\tilde{\chi}^+_1\to\tilde{\chi}^0_1e^+\nu_e$ proceeds via
$\tilde{e}_L$, $\tilde{\nu}_e$ and $W$ exchange so that the $M_1$
dependence of the cross section and of $A_{\mathrm{FB}}$ is determined 
by the interplay between the $M_1$ dependence of the LSP mass and the
$M_1$ dependence and the relative magnitude of the relevant couplings.
Similar as for the $m_{\tilde{\nu}_e}$ dependence the $M_1$ dependence 
of the cross section reflects that of the leptonic branching ratio of
the chargino decay.
For large $M_1$ the mass of the
LSP and the couplings are nearly independent of $M_1$ \cite{claus},
leading to 
the flat run of the cross section and the forward-backward
asymmetry. 
Between $M_1=50$ GeV and $M_1=100$ GeV the cross section increases
approximately 10\% and for $M_1>100$ GeV it is nearly independent of
$M_1$. The forward-backward asymmetry, however, shows a pronounced
$M_1$-dependence between $M_1=50$ GeV and $M_1=150$ GeV.
Since there are no ambiguities it should be possible to
determine $M_1$ in this region from a
measurement of the forward-backward asymmetry.

\section{Conclusion}
We have studied the production of charginos at a 500 GeV Linear
Collider in the $\gamma\gamma$ mode including the complete spin
correlations between production and the subsequent leptonic decay. 
For $m_{\tilde{\nu}_e}< 250$ the cross section is strongly dependent
on $m_{\tilde{\nu}_e}$. The 
forward-backward asymmetry of the decay leptons shows a pronounced
dependence on the gaugino mass parameter $M_1$ in the region
50 GeV$<M_1<150$ GeV and on the sneutrino mass for
$m_{\tilde{\nu}_e}\stackrel{\textstyle<}{\sim}400$ GeV.
Since neither the $M_1$ dependence nor the
$m_{\tilde{\nu}_e}$ dependence of the cross section and the asymmetry
show ambiguities, this process should allow to test the GUT relation
between $M_1$ and $M_2$ and to determine the sneutrino mass.

\section*{Acknowledgements}
We are grateful to Gudi Moortgat-Pick and Claus Bl\"ochinger for
numerous valuable discussions. We also
thank Fabian Franke for many helpful comments
on the manuscript. This work was supported by the Deutsche
Forschungsgemeinschaft under contract no. FR 1064/4-1 and the
Bundesministerium f\"ur Bildung und Forschung (BMBF) under contract
no. 05 HT9WWA 9.

\end{document}